\documentclass[useAMS,usenatbib,usegraphicx]{mn2e}
\usepackage{times}
\usepackage{amsmath}
\usepackage{amssymb}

\input{psfig.sty}

\newcommand{\AIPS}{{$\cal AIPS\/$}}

\def\l1.4{$L_{\rm 1.4GHz}$}
\def\s1.4{$S_{\rm 1.4GHz}$}

\def\kms{km\,s$^{-1}$}

\def\Tb{$T_{\rm b}$}
\def\K{{\sc k}}

\def\gs{\mathrel{\raise0.35ex\hbox{$\scriptstyle >$}\kern-0.6em
\lower0.40ex\hbox{{$\scriptstyle \sim$}}}}
\def\ls{\mathrel{\raise0.35ex\hbox{$\scriptstyle <$}\kern-0.6em
\lower0.40ex\hbox{{$\scriptstyle \sim$}}}}
\def\m@th{\mathsurround=0pt }
\def\eqalign#1{\null\,\vcenter{\openup1\jot \m@th
 \ialign{\strut\hfil$\displaystyle{##}$&$\displaystyle{{}##}$\hfil
 \crcr#1\crcr}}\,}

\title[EVLA imaging of CO(1--0) in submillimetre galaxies]
      {Tracing the molecular gas in distant submillimetre galaxies
       via CO(1--0) imaging with the EVLA}

     \author[Ivison et al.]  {R.\,J.\ Ivison,$^{\! 1,2}$
       P.\,P.~Papadopoulos,$^{\! 3}$ Ian Smail,$^{\! 4}$
       T.\,R.~Greve,$^{\! 5}$ A.\,P.\ Thomson,$^{\! 2}$ \and E.\,M.\
       Xilouris$^6$ and S.\,C.~Chapman$^7$
       \vspace*{1mm}\\
       $^1$ UK Astronomy Technology Centre, Science and Technology
       Facilities Council,
       Royal Observatory, Blackford Hill, Edinburgh EH9 3HJ\\
       $^2$ Institute for Astronomy, University of Edinburgh,
       Blackford Hill,
       Edinburgh EH9 3HJ\\
       $^3$ Argelander-Instit\"ut f\"ur Astronomie, Auf dem H\"ugel
       71,
       D-53121, Germany\\
       $^4$ Institute for Computational Cosmology, Durham University,
       South Road, Durham DH1 3LE\\
       $^5$ Dark Cosmology Centre, Niels Bohr Institute, University of
       Copenhagen, Juliane Maries Vej 30, 2100 Copenhagen \O, Denmark\\
       $^6$ Institute of Astronomy and Astrophysics, National
       Observatory of Athens, I.\ Metaxa and Vas.\ Pavlou Streets, P.\
       Penteli,
       GR-15236 Athens, Greece\\
       $^7$ Institute of Astronomy, University of Casmbridge,
       Madingley Road, Cambridge CB3 0HA}

\date{Accepted 2010 November 15. Received 2010 November 2; in original form 2010 September 2}

\pagerange{000--000}

\begin{document}

\maketitle

\begin{abstract}
  We report the results of a pilot study with the Expanded Very Large
  Array (EVLA) of $^{12}$CO $J\!=\!1\!-\!0$ emission from four
  submillimetre-selected galaxies (SMGs) at $z= 2.2$--2.5, each with
  an existing detection of $^{12}$CO $J\!=\!3\!-\!2$, one of which
  comprises two distinct spatial components. Using the EVLA's most
  compact configuration we detect strong, broad (medians: 990\,\kms\
  {\sc fwzi}; 540\,\kms\ {\sc fwhm}) $J\!=\!1\!-\!0$ line emission
  from all of our targets -- coincident in position and velocity with
  their $J\!=\!3\!-\!2$ emission. The median line width ratio,
  $\sigma_{\rm 1-0}/\sigma_{\rm 3-2}=1.15\pm 0.06$, suggests that the
  $J\!=\!1\!-\!0$ is more spatially extended than the $J\!=\!3\!-\!2$
  emission, a situation confirmed by our maps which reveal velocity
  structure in several cases and typical sizes of $\sim$16\,kpc {\sc
    fwhm}. The median brightness temperature (\Tb) ratio is
  $r_{3-2/1-0}=0.55\pm0.05$, consistent with local galaxies with
  $L_{\rm IR}>10^{11}$\,L$_{\odot}$, noting that our value may be
  biased high because of the $J\!=\!3\!-\!2$-based sample
  selection. Naively, this suggests gas masses roughly 2$\times$
  higher than estimates made using higher-$J$ transitions of CO, with
  the discrepency due entirely to the difference in assumed \Tb\
  ratio. We also estimate molecular gas masses using the $^{12}$CO
  $J\!=\!1\!-\!0$ line and the observed global \Tb\ ratios, assuming
  standard underlying \Tb\ ratios for the non-star-forming and
  star-forming gas phases as well as a limiting star-formation
  efficiency (SFE) for the latter in all systems, i.e.\ without
  calling upon $X_{\rm CO}$ ($\equiv\alpha$). Using this new method,
  we find a median molecular gas mass of $(2.5\pm 0.8) \times
  10^{10}$\,M$_{\odot}$, with a plausible range stretching up to
  3$\times$ higher. Even larger masses cannot be ruled out, but are
  not favoured by dynamical constraints: the median dynamical mass
  within $R\sim 7$\,kpc for our sample is $(2.3\pm1.4) \times
  10^{11}$\,M$_{\odot}$, or $\sim$6$\times$ more massive than
  UV-selected galaxies at this epoch. We examine the Schmidt-Kennicutt
  (S-K) relation for all the distant galaxy populations for which CO
  $J\!=\!1\!-\!0$ or $J\!=\!2\!-\!1$ data are available, finding small
  systematic differences between galaxy populations. These have
  previously been interpreted as evidence for different modes of star
  formation, but we argue that these differences are to be expected,
  given the still considerable uncertainties, certainly when
  considering the probable excitation biases due to the molecular
  lines used, and the possibility of sustained S-K offsets during the
  evolution of individual gas-rich systems. Finally, we discuss the
  morass of degeneracies surrounding molecular gas mass estimates, the
  possibilities for breaking them, and the future prospects for
  imaging and studying cold, quiescent molecular gas at high
  redshifts.
\end{abstract}

\begin{keywords}
  galaxies: evolution --- galaxies: high-redshift ---
  galaxies: starburst ---
  infrared: galaxies --- radio lines: galaxies
\end{keywords}

\section{Introduction}

The star-formation density contributed by ultraluminous infrared
galaxies (ULIRGs) appears to increase out to $z\gs 2$
\citep[e.g.][]{chapman05, wardlow10}.  These galaxies are often
heavily obscured by dust and hence the bulk of their luminosity is
radiated in the rest-frame far-infrared (far-IR) and observed in the
submillimetre (submm) band, hence their epitette: `submm galaxies'
(SMGs). SMGs have the potential to form the stellar mass of an
L$^\ast$ galaxy in a single event \citep[e.g.][]{lilly99, smail04,
  swinbank06, hainline10}. To accomplish this feat, however, SMGs must
have sufficiently large reservoirs of cold gas,
$\gs$10$^{11}$\,M$_{\odot}$. The first searches for molecular gas
emission, via the $J\!=\!3\!-\!2$ or $4\!-\!3$ transitions of
$^{12}$CO, were successful in detecting significant quantities of gas
in several SMGs \citep{frayer98, frayer99, ivison01}. Subsequent,
larger surveys with the Plateau de Bure Interferometer (PdBI) greatly
expanded this work \citep[e.g.][]{downes03, genzel03, kneib05,
  greve05, tacconi06, bothwell10}.

Observations of the molecular gas within high-redshift galaxies
provides powerful insight into the physics of star formation in these
systems \citep{solomon05}, and allows comparisons with local
systems. More specifically they allow us to: a) probe the mass and
extent of the reservoir of molecular gas available for fueling their
prodigious starbursts; b) determine the dynamical mass of the host
galaxy, free from the uncertainties arising from outflows and patchy
dust extinction which plagues optical and near-IR spectroscopic
studies \citep[e.g.][]{swinbank06, ivison10leblob}; c) derive their
gas-mass fraction, $M_{\rm gas}({\rm H}_2)/M_{\rm dyn}$. The dynamical
mass is a strong indicator of the `end' state of such systems in the
present Universe while a well-determined gas-mass fraction indicates
their likely evolutionary status at the look-back time where they are
observed. Finally, gas-consumption timescales, $\langle \tau_{\rm gas}
\rangle = M({\rm H}_2)/{\rm SFR}$, where SFR is the star-formation
rate, gives the minimum timescale for the conclusion of their
star-forming (SF) episodes.

While the PdBI studies of SMGs represent a considerable advance, they
are fundamentally limited by their focus on high-$J$ $^{12}$CO lines
whose high excitation requirements ($n_{\rm crit}\sim
10^4$--$10^5$\,cm$^{-3}$, $E_{\rm u}/k_{\rm B}\sim 50$--150\,\K)
confine the emission from such transitions to regions of active star
formation, rather than tracing the total available reservoir of gas
within a galaxy.  In the nearby archetypal starburst, M\,82, such
lines would reveal only the highly excited, SF molecular gas in its
inner 400\,pc \citep{mao00, weiss01} rather than the more massive,
low-excitation gas component that extends $\sim 1.7$\,kpc beyond its
centre \citep{walter02, weiss05m82}. This also suggests potential
spatial biases in the high-$J$ $^{12}$CO emission -- which exhibit
typical half-light radii of 0.8--2.8\,kpc for SMGs \citep{tacconi08}
-- means they may not trace the true dimensions (or kinematics) of the
total molecular gas distribution.

A recent comparison of $^{12}$CO line ratios between local,
IR-luminous galaxies and distant SMGs has provided strong indications
that several high-redshift systems must contain significant amounts of
colder, non-SF gas or -- for a few compact, extreme starbursts -- that
they may suffer high optical depths at short submm wavelengths due to
dust \citep{papadopoulos10a}. Thus their high-$J$ $^{12}$CO line
emission may not be a good tracer of the total CO-rich molecular gas
mass, its distribution, or the total enclosed dynamical masses, as is
usually assumed \citep[e.g.][]{tacconi06}.  Indeed the low
brightness-temperature (\Tb) ratios of $^{12}$CO $J\!=\!7\!-\!6$ (or
$6\!-\!5$) to $3\!-\!2$ ($r_{7-6/3-2}\la 0.3$) measured in several
SMGs \citep{tacconi06} {\it are not} typical of dense, warm, SF gas
providing strong but still circumstantial evidence that high-$J$
$^{12}$CO line studies may miss a critical gas component in SMGs.

The earliest evidence of low, Milky Way-type, global CO line
excitation in a distant SF system was uncovered in the submm-bright
extremely red object, HR\,10 \citep{pi02}. Since then a handful of
SMGs have been observed in $^{12}$CO $J\!=\!1\!-\!0$ \citep{greve03,
  hainline06, ivison10leblob, swinbank10, carilli10, harris10,
  frayer10} and several show evidence of substantial reservoirs of
$^{12}$CO $J\!=\!1\!-\!0$ with $r_{\rm 3-2/1-0}\sim 0.5$ \citep[see
also][]{danielson10}. A systematic under-estimate of molecular gas
mass via high-$J$ CO line emission could lie at the heart of the
apparent discrepancy between the gas-depletion timescales for SMGs,
40--100\,Myr, based on high-$J$ observations \citep{greve05}, and the
proposed lifetimes of these luminous starbursts, $\sim$300\,Myr
\citep{swinbank06, swinbank08}, though strong feedback events will
punctuate the evolutionary path of any starburst and these are
expected to lengthen any putative gas-consumption timescales.  This,
in turn, influences our understanding of the evolutionary links
between SMGs and other high-redshift populations, e.g.\ quasars or
passive galaxies \citep{chapman05, wardlow10} which rely on duty-cycle
and space-density arguments, which then influences our interpretation
of their likely descendants at $z\sim 0$ \citep{swinbank08}.
Similarly, the spatial bias in the high-$J$ $^{12}$CO dynamical
tracers employed for SMGs means we might be missing important
signatures in the velocity fields at large radii, perhaps reflecting
disk-like rotation or even evidence for cold-flow accretion
\citep[e.g.][]{dekel09b}.

%
%
\begin{table*}
  \centering
  \caption{SMG sample and observing log.\label{tab:sample}}
  \begin{tabular}{lcccll}
    \hline \hline
    \multicolumn{1}{l}{Target name} &
    \multicolumn{1}{c}{$z_{\rm CO3-2}^a$} &
    \multicolumn{1}{c}{Deboosted} &
    \multicolumn{1}{c}{$S_{\rm 1.4GHz}$} &
    \multicolumn{1}{c}{Observing dates (2010)} &
    \multicolumn{1}{l}{Alternative target names} \\
    \multicolumn{1}{l}{SMM\,J...} &
    \multicolumn{1}{c}{} &
    \multicolumn{1}{c}{$S_{\rm 850\mu m}$ (mJy)} &
    \multicolumn{1}{c}{($\mu$Jy)} &
    \multicolumn{1}{c}{(2.0--2.1\,hr per track, on-source)} &
    \multicolumn{1}{l}{} \\
    \hline
123549.44+621536.8    &2.202&$8.3\pm2.5$&$81\pm5$   & May 2, 14, 15, 16, Jun 20, 27, Jul 6& HDF\,76 \\ 
123707.21+621408.1-NE &2.486&$10.7\pm2.7$&$39\pm8$  & Apr 12, 14, 22, May 4, 6&HDF\,242, GN\,19\\ 
123707.21+621408.1-SW &     &            &$30\pm8$  & &\\
163650.43+405734.5    &2.385&$8.2\pm1.7$&$242\pm11$ & Apr 16, 19 ($\times 2$), 20, May 8, 11, 12& N2\,850.4, N2\,1200.10\\ 
163658.19+410523.8    &2.452&$10.7\pm2.0$&$115\pm11$& Apr 22, 27, May 8, 14, Jun 20, 21, Jul 7&  N2\,850.2\\ 
    \hline
  \end{tabular}

{\small
Notes: $^a$ CO $J=3-2$ redshifts from \citet{tacconi06};
for SMM\,J123707, we quote the value used to determine
velocities in that work.}
\end{table*}

To address these concerns we have undertaken a pilot study with the
National Radio Astronomy Observatory's (NRAO's\footnotemark[1])
Expanded Very Large Array (EVLA) of $^{12}$CO $J\!=\!1\!-\!0$ emission
from four well-studied SMGs, with $S_{\rm 850\mu m} \sim 10$\,mJy at
$z =2.2$--2.5, two each in the Great Observatories Origins Surveys
(GOODS) North and European Large Area Infrared Survey (ELAIS) N2 submm
survey fields \citep{scott02, borys03}, pinpointed accurately via
arcsec-resolution radio continuum imaging \citep{ivison02, chapman05},
with redshifts determined via Keck spectroscopy \citep{chapman03,
  chapman05} and confirmed via detections of $^{12}$CO $J\!=\!3\!-\!2$
at PdBI \citep{neri03, greve05, tacconi06, tacconi08}. Ultimately, the
combination of EVLA and PdBI imaging will allows us to compare the gas
masses, morphologies and dynamics derived from the $^{12}$CO
$J\!=\!1\!-\!0$ and $3\!-\!2$ or higher-$J$ lines in these intense
starbursts. In this first paper we explore those parameters probed by
the initial phase of our EVLA survey, conducted at relatively low
resolution using the EVLA's most compact configuration during
shared-risk time.

Throughout the paper we use a cosmology with $H_0 =
71$\,km\,s$^{-1}$\,Mpc$^{-1}$, $\Omega_m=0.27$, $\Omega_\Lambda =
0.73$ which gives a median angular scale of 8.2\,kpc\,arcsec$^{-1}$
for our sample.

\footnotetext[1]{NRAO is operated by Associated Universities Inc., under 
a cooperative agreement with the National Science Foundation.}

\section{Observations and Data Reduction}
\label{observations}

Our sample was chosen such that the $^{12}$CO $J\!=\!1\!-\!0$ line is
redshifted to the $\sim$33--36\,GHz frequency range, where both
sub-band pairs of the EVLA's new Ka-band receivers can be utilised,
yielding up to 256\,MHz of instantaeous bandwidth during the earliest
shared-risk phase of EVLA commissioning with the new Wideband
Interferometric Digital Architecture (WIDAR) correlator (see
Table~\ref{tab:sample}). We overlapped the two $64\times 2$-MHz
dual-polarisation sub-bands by a total of 10 channels -- to minimise
issues with the edge channels -- and we centred the lines 36\,MHz to
the red, using the redshifts published by \citet{tacconi06,
  tacconi08}. This latter precaution proved unnecessary. Our approach
ultimately yielded a single 236-MHz dataset for each target,
$\sim$2,000\,\kms\ of coverage with 16.7--18.2-\kms\ channels,
covering $uv \sim 5$--$100\rm k\lambda$.

Each target was observed for between five and seven 3-hr tracks during
2010 April--July (Table~\ref{tab:sample}), a period during which
$\sim$15 functional Ka-band receivers were typically available and the
EVLA was in its most compact configuration (D), recording data every
1\,s, with a resultant data rate of $\sim$6.3\,GB\,hr$^{-1}$
[programme AS1013]. The tracks were scheduled flexibly to ensure
appropriate weather conditions for these high-frequency
observations. After 2010 May 12, several antennas were plagued by
phase jumps, but most of these data were salvaged. Three tracks
suffered severe phase jumps or unsuitable weather and were discarded.
Antenna pointing was checked every 90\,min at 5\,GHz. Each track
contained a 5-min scan of 1331+305 (3C\,286) for absolute flux
calibration (1.87\,Jy at 35.1\,GHz), and regular (every $\ls$5\,min)
scans of the bright, local calibrators, 1302+575 and 1642+394
($\sim$0.35 and $\sim$7.0\,Jy, respectively). The latter has been
monitored at 33.75\,GHz \citep{davies09} and we found consistent flux
densities when bootstrapping from 0137+331, albeit with evidence of
the variability found by \citet{franzen09}.

Editing and calibration were accomplished within \AIPS\ ({\sc
  31dec10}). For 1331+305, we used an appropriately scaled 22.5-GHz
model to determine gain solutions; the other calibrators were used to
determine the spectral variation of the gain solutions (the
`bandpass'), after first removing atmospheric phase drifts on a
timescale of 6--12\,s via self-calibration with a simple point-source
model. Despite the significant increase in data volume relative to the
old VLA correlator, we were able to employ standard \AIPS\ recipes
throughout the data-reduction process.

\section{Results}

\subsection{Infrared luminosities}
\label{lir}

On several occasions in the subsequent discussion, e.g.\
\S\ref{twophases}, \ref{tau} and \ref{s-k}, we shall call upon SFRs
determined from rest-frame 8--1,000-$\mu$m luminosities, $L_{\rm IR}$,
so we start by describing these measurements.

For our SMGs, plus 13 others available in the literature with
measurements of $^{12}$CO $J\!=\!1\!-\!0$ or $J\!=\!2\!-\!1$ and
comparable selection biases and general properties \citep{greve03,
  greve05, hainline06, frayer08, schinnerer08, knudsen09,
  ivison10leblob, ivison10fts, harris10, carilli10, swinbank10}, we
derive $L_{\rm IR}$ by fitting the spectral energy distribution (SED)
template from \citet{pope08} and that of SMM\,J2135$-$0102
\citep{swinbank10,ivison10fts} to the available photometric data,
ignoring the radio photometry for those sources known to suffer
contamination by AGN: SMM\,J163650 \citep{smail03}, SMM\,J02399
\citep{ivison10leblob} and SMM\,J14009 \citep{ivison00, weiss09}. The
SMM\,J2135 SED gave better fits and we adopt these values, which are
10--20 per cent higher than those given by the \citet{pope08}
template. We then convert these into SFRs, following
\citet{kennicutt98a}. Our five SMGs have a mean $L_{\rm IR}$ of
$(5.4\pm0.8)\times 10^{12}$\,L$_{\odot}$ (see Table~\ref{tab:masses})
and a corresponding SFR of $930\pm 140$\,M$_{\odot}$\,yr$^{-1}$ (for
the full sample, $L_{\rm IR} =(5.5\pm 0.9)\times 10^{12}$\,L$_{\odot}$
and $950\pm 155$\,M$_{\odot}$\,yr$^{-1}$).

\subsection{Spectra and morphologies}
\label{specmorph}

\begin{figure*}
\centerline{\psfig{figure=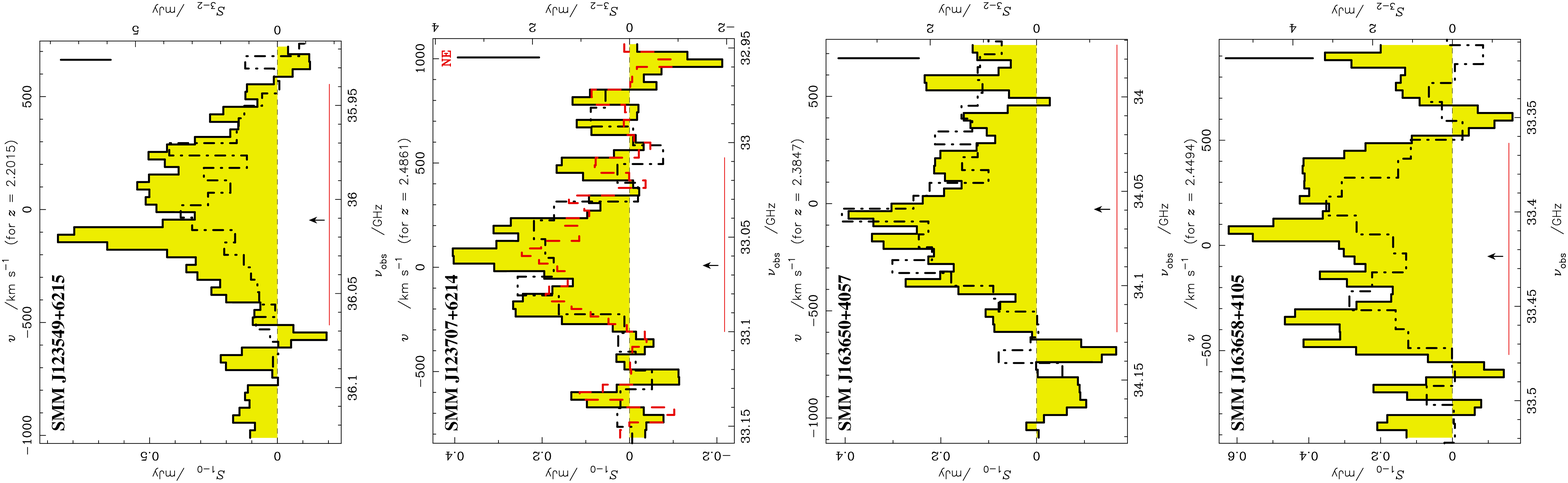,width=2.41in,angle=270} \hspace*{2cm}
\psfig{figure=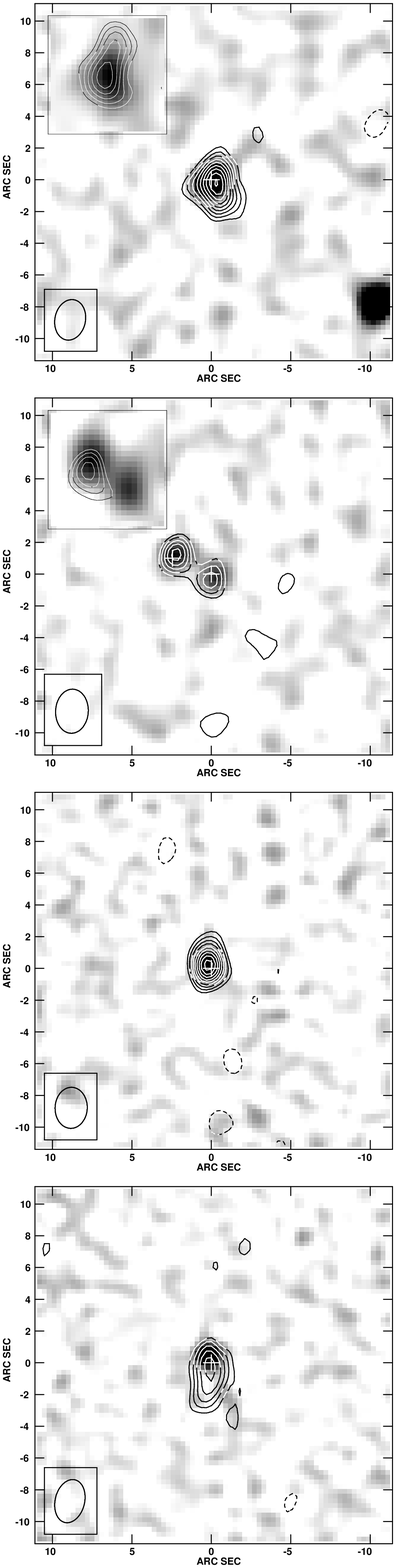,width=2.03in,angle=0}}
\caption{CO $J\!=\!1\!-\!0$ spectra and lightly {\sc clean}ed images for the
  SMGs in our study.  On the left we show the spectra, integrated over
  a region of $\sim$3 beam areas in the datacubes, with the
  corresponding PdBI CO $J\!=\!3\!-\!2$ spectra from \citet{tacconi06}
  shown as dot-dashed lines, scaled by $9^{-1}\times$ to be on the
  same Rayleigh-Jeans \Tb\ scale. Arrows indicate the
  CO $J\!=\!3\!-\!2$ line centres. The CO $J\!=\!1\!-\!0$ spectrum of the NE
  component of SMM\,J123707 is shown in red. All our targets are
  well-detected with CO $J\!=\!1\!-\!0$ {\sc fwhm} of
  $\sim$300--800\,km\,s$^{-1}$,
  marginally broader than the corresponding CO $J\!=\!3\!-\!2$
  emission. The frequency range used to create the images in the
  right-hand panels are indicated below each spectrum.  A typical
  $\pm$1-$\sigma$ error bar is shown at the top right of each
  panel. Velocities are relative to the CO $J\!=\!1\!-\!0$ redshifts
  listed in Table~\ref{tab:properties} and the spectra have been
  Hanning smoothed with an four-channel triangular kernel with 4-MHz
  ($\sim$35-km\,s$^{-1}$) spectral resolution. {\it Right:} CO
  $J\!=\!1\!-\!0$ emission
  integrated over the channels indicated in the corresponding
  spectrum, displayed as contours ($-3$, 3, 4... $\times\sigma$ where
  $\sigma$ is typically $\sim$20\,$\mu$Jy\,beam$^{-1}$), superimposed
  on linear greyscales of the 1.4-GHz continuum emission.  We see that
  all targets are marginally or well resolved at the resolution of our
  map, with SMM\,J123707 comprising two components. The beam is shown
  in the bottom left corner of each image. Insets show 6- by
  6-arcsec postage stamps of CO $J\!=\!1\!-\!0$ for SMM\,J123549 (a
  220-\kms-wide bin centred at $-150$\,\kms, displayed as contours,
  superimposed on a greyscale of the total CO
  $J\!=\!1\!-\!0$ emission) and SMM\,J123707 (contiguous 390-\kms-wide
  blue and red bins represented by the contours and the underlying greyscale,
  respectively, illustrating the velocity gradient across SMM\,J123707-NE).}
  \label{fig:both}
\end{figure*}

We construct maps of each source, stepping through the frequency range
of the data, examining image cubes with velocity resolutions ranging
from 2--16 channels (35--250\,\kms). The synthesised beam for a
natural weighting scheme is typically 2.7 by 2.0\,arcsec with a
position angle (PA) near $-5^{\circ}$, or 3.2 by 2.6\,arcsec when
employing a Gaussian taper that reaches 30 per cent at
80\,k$\lambda$. We find strong $^{12}$CO $J\!=\!1\!-\!0$ line emission
from each of the four SMGs, at the expected positions and frequencies
derived from the $^{12}$CO $J\!=\!3\!-\!2$ emission, though we note
that some published redshifts/spectra are misleading -- e.g.\ for
SMM\,J123707, \citet{tacconi06} quote $z=2.490$ rather than
$z=2.486$. We see two distinct spatial components in SMM\,J123707, so
we hereafter refer to a sample of five SMGs using the suffixes -NE and
-SW for SMM\,J123707, following \citet{tacconi06}. We extract a
spectrum for the emission in an area corresponding to around three
synthesised beams, centred on each of our targets (SMM\,J123707-NW and
-SW are combined in this case) and show these in Fig.~\ref{fig:both},
after Hanning smoothing with a four-channel triangular kernel. All
sources display broad lines, with full widths at zero intensity of
$\sim$1,000\,km\,s$^{-1}$. We measure the line fluxes by integrating
the emission across the velocity ranges reported in
Table~\ref{tab:sample} (no continuum correction is applied as we
detect no significant continuum emission from any of the SMGs,
$3\sigma < 60$\,$\mu$Jy, as expected).  We assess the uncertainty in
the velocity-integrated fluxes from the variance of the off-line
emission and we report the flux and associated error in
Table~\ref{tab:properties}. We also list the CO $J\!=\!3\!-\!2$ line
intensities \citep[from][]{tacconi08} for our five SMGs, having
checked for consistency with the low-resolution fluxes of
\citet{greve05}.

%
%

We calculate the intensity-weighted redshift for the emission in the
velocity window and report this in Table~\ref{tab:properties}.  It is
clear from \citet{tacconi08} that some of the $^{12}$CO
$J\!=\!3\!-\!2$ lines are better fit by two Gaussian profiles and this
is also true for several of the $^{12}$CO $J\!=\!1\!-\!0$ lines;
however, for the sake of uniformity we determine the
intensity-weighted second moment of the velocity distribution and
report this as the equivalent 1-$\sigma$ width of a Gaussian with the
same second moment in Table~\ref{tab:properties}. We find a median
{\sc fwhm} of $540\pm 110$\,km\,s$^{-1}$ for our target SMGs, which is
nearly double the median {\sc fwhm} of $300\pm 20$\,\kms\ measured for
the $^{12}$CO $J\!=\!1\!-\!0$ emission in ULIRGs by
\cite{solomon97}. For comparison, we also calculate the equivalent
1-$\sigma$ width of the $^{12}$CO $J\!=\!3\!-\!2$ emission in the same
velocity range from the spectra shown in \citet{tacconi08}, finding a
median ratio between the two line widths of $\sigma_{\rm
  1-0}/\sigma_{\rm 3-2}=1.15\pm 0.06$.  Assuming that the dynamics of
the central regions of these systems are dominated by baryonic gas
disks then this suggests that the $^{12}$CO $J\!=\!1\!-\!0$ is more
spatially extended than the $^{12}$CO $J\!=\!3\!-\!2$ emission, a
situation confirmed by our maps.

Fig.~\ref{fig:both} also shows the maps of the $^{12}$CO
$J\!=\!1\!-\!0$ emission.  We mark the centroids of the $^{12}$CO
$J\!=\!3\!-\!2$ emission for all of the sources, confirming that the
$^{12}$CO $J\!=\!1\!-\!0$ emission is spatially coincident in all
cases, although we note that in SMM\,J163658 there is evidence that
the 1.4-GHz continuum is offset to the north of the $J\!=\!3\!-\!2$
emission which is, in turn, offset to the north of the $J\!=\!1\!-\!0$
centroid -- a situation reminiscent of the complex AGN/starburst,
SMM\,J02399$-$0136 \citep{ivison10leblob}.

\footnotetext[2]{The ELAIS~N2 continuum image was recreated from the
  raw data [programmes AI91 and AD432] to avoid the astrometric issue
  described by \citet{morrison10}. See also {\tt
    http://science.nrao.edu/evla/archive/issues/\#010}.}

Even though these $^{12}$CO $J\!=\!1\!-\!0$ maps were taken with the
most compact EVLA configuration, they resolve all of the systems
except SMM\,J123707-SW, one of two $^{12}$CO $J\!=\!1\!-\!0$ emitters
seperated by $\sim$3\,arcsec ($\sim$23\,kpc) in the field of
SMM\,J123707, as previously identified in $^{12}$CO $J\!=\!3\!-\!2$
with PdBI and at 1.4\,GHz with the VLA by \citet{tacconi06}. The
north-eastern component within SMM\,J123707 and the other three SMGs
have typical {\sc fwhm} of $\sim$2\,arcsec or $\sim$16\,kpc.  These
sizes are larger than the 0.8--2.8\,kpc half-light radii deduced from
the higher-$J$ transitions by \citet{tacconi08}, as expected from the
larger line widths.  Combining our data with the forthcoming
C-configuration EVLA observations of $^{12}$CO $J\!=\!1\!-\!0$ will
ensure we have sufficient sensitivity and resolution to confirm this
suggestion.

We also find evidence for velocity structure within several of the
sources. In SMM\,J123549 we identify a northern spur to the
$J\!=\!1\!-\!0$ emission (see inset, Fig.~\ref{fig:both}), centred at
$-150$\,\kms, which is not seen in $J\!=\!3\!-\!2$. This feature is
coincident with a prominent, narrow feature in the $J\!=\!1\!-\!0$
spectrum and suggests that the system may contain two components
\citep[cf.\ 4C\,60.07 --][]{ivison08}. We find a velocity difference
of just $80\pm 60$\,km\,s$^{-1}$ between the two components of
SMM\,J123707, suggesting that they are orbiting one another in the
plane of the sky with a physical separation of $\sim$25\,kpc. We also
see velocity structure within SMM\,J123549 and most strikingly a
velocity gradient of $\sim$600\,km\,s$^{-1}$ across SMM\,J123707-NE
over a spatial scale of $\sim$2\,arcsec ($\sim$16\,kpc -- see inset,
Fig.~\ref{fig:both}), though our spatial resolution is currently
insufficient to determine whether the velocity structure is well
ordered.

\subsection{Brightness temperature ratios: average gas excitation }
\label{tbratios}

Combining our new $^{12}$CO $J\!=\!1\!-\!0$ data with the $^{12}$CO
$J\!=\!3\!-\!2$ luminosities from \citet{tacconi08}, we derive a
weighted-mean \Tb\ ratio of $r_{\rm 3-2/1-0}=0.56\pm 0.05$ (the median
ratio and bootstrapped error are $0.55\pm 0.05$).  Including similar
observations of the same transitions in the five $z\sim 2$ SMGs from
\citet{ivison10leblob}, \citet{harris10} and \citet{swinbank10}, we
derive a weighted-mean \Tb\ ratio of $r_{3-2/1-0}=0.65\pm 0.02$ and a
median of $0.58\pm 0.05$.  The distributions of $r_{\rm 3-2/1-0}$ for
our sample and the literature sources are statistically
indistinguishable.

The weighted-mean \Tb\ ratio for the \citet{yao03} sub-sample of local
IR-luminous galaxies with rest-frame 8--1,000-$\mu$m luminosities,
$L_{\rm IR}\geq 10^{11}$\,L$_{\odot}$, is $r_{\rm 3-2/1-0}=0.57\pm
0.06$ with a median of $0.63\pm 0.10$, similar to that of
high-redshift SMGs \citep[see also][]{harris10}.  This is also
consistent with the average excitation conditions of local LIRGs
studied by \citeauthor{yao03} but $L_{\rm IR}$ and the molecular gas
mass is typically an order of magnitude larger for the SMGs.  This
similarity, while reassuring, does not mean that the physical state of
the molecular gas in SMGs and LIRGs is identical. Indeed, $r_{\rm
  3-2/1-0}$ is often low for the LIRGs studied by \citeauthor{yao03},
which is atypical of dense, SF molecular gas (where this ratio would
approach unity); this is thought to be caused by a diffuse, unbound,
warm gas phase that is frequently found in galactic nuclei. The
physical scale sampled by our beam in SMGs at $z\sim 2.4$
($\sim$20\,kpc) is larger than that employed for the LIRGs studied by
\citeauthor{yao03} ($\sim$2--6\,kpc) and may sample an extended, cold
molecular gas phase that can suppress the global $r_{\rm 3-2/1-0}$ (an
issue discussed in detail in \S\ref{twophases}).

We highlight two features of our $r_{\rm 3-2/1-0}$ distribution.
Firstly, we are using what is effectively an $^{12}$CO
$J\!=\!3\!-\!2$-selected sample for this analysis and it is possible
that SMGs with lower $r_{\rm 3-2/1-0}$ ratios remained undetected by
the $^{12}$CO $J\!=\!3\!-\!2$ surveys and were thereby excluded from
our study.  We note that a programme to search for $^{12}$CO
$J\!=\!1\!-\!0$ emission using the Zpectrometer instrument on the
Green Bank Telescope (GBT) from SMGs which were undetected in
$^{12}$CO $J\!=\!3\!-\!2$ by PdBI has been terminated due to problems
with GBT's Ka receiver.  Secondly, we stress that there is a wide
range in the apparent $r_{\rm 3-2/1-0}$ values within the SMG sample
but that formally these are consistent with our error-weighted mean
value, given the quoted uncertainties, so we have yet to uncover the
variation in $r_{\rm 3-2/1-0}$ seen locally \citep{yao03}.

%
%
\begin{table*}
  \centering
  \caption{SMG observed properties.\label{tab:properties}}
  \begin{tabular}{lcccccccc}
    \hline 
    \multicolumn{1}{l}{Name} &
    \multicolumn{1}{c}{$z_{\rm CO(1-0)}$} &
    \multicolumn{1}{c}{$I_{\rm 1-0}$} &
    \multicolumn{1}{c}{$I_{\rm 3-2}$} &
    \multicolumn{1}{c}{$r_{\rm 3-2/1-0}$} &
    \multicolumn{1}{c}{$r_{\rm 6-5/3-2}$} &
    \multicolumn{1}{c}{$\sigma_{\rm 1-0}^a$} &
    \multicolumn{1}{c}{{\sc fwzi}} &
    \multicolumn{1}{c}{$R$} \\
    \multicolumn{1}{c}{} &
    \multicolumn{1}{c}{} &
    \multicolumn{1}{c}{(Jy\,km\,s$^{-1}$)} &
    \multicolumn{1}{c}{(Jy\,km\,s$^{-1}$)} &
    \multicolumn{1}{c}{} & 
    \multicolumn{1}{c}{} & 
    \multicolumn{1}{c}{(km\,s$^{-1}$)} &
    \multicolumn{1}{c}{(km\,s$^{-1}$)} &
    \multicolumn{1}{c}{(kpc)} \\
    \hline
SMM\,J123549   &$2.2015\pm 0.0002$&$0.32\pm0.04$&$1.6\pm0.2$  &$0.56\pm 0.10$&$0.36\pm 0.08$ & $230\pm 20$&1050&$7\pm 2$\\
SMM\,J123707-NE&$2.4870\pm 0.0005$&$0.09\pm0.02$&$0.32\pm0.09$&$0.40\pm 0.14$&...            & $200\pm 35$&830 & $7\pm 2$\\
SMM\,J123707-SW&$2.4861\pm 0.0004$&$0.12\pm0.02$&$0.59\pm0.09$&$0.55\pm 0.12$&...            & $140\pm 30$&920 & $<3$    \\
SMM\,J163650   &$2.3847\pm 0.0004$&$0.34\pm0.04$&$2.3\pm0.3$  &$0.75\pm 0.13$&$0.23\pm 0.05$ & $330\pm 35$&1340&$6\pm 2$ \\
SMM\,J163658   &$2.4494\pm 0.0002$&$0.37\pm0.07$&$1.8\pm0.2$  &$0.54\pm 0.12$&$0.33\pm0.07$  & $295\pm 10$&990 &$11\pm 3$\\
    \hline
  \end{tabular}

{\small
Note: $^a$ {\sc fwhm} = $2 \sqrt{2 \ln 2} \times \sigma$.}
\end{table*}

\subsection{The plausible range of molecular gas masses}
\label{masses}

The key advantage of observing the lowest $^{12}$CO $J\!=\!1\!-\!0$
transition is that it allows us to determine their molecular gas mass
and its distribution in a manner identical to that used in the local
Universe, which permits a direct comparison of these systems to local
IR-luminous galaxies. A firm lower limit on the molecular gas mass can
be obtained by assuming local thermodynamic equilibrium and an
optically thin transition where:

\begin{eqnarray}
\frac{M({\rm H}_2)}{L'_{\rm CO1-0}}=X^{\rm thin}_{\rm CO} \sim & 0.08 
\left[\frac{g_1}{Z} \, e^{-T_{\circ}/T_{\rm k}}
\left(\frac{J(T_{\rm k})-J(T_{\rm bg})}{J(T_{\rm k})}\right)\right]^{-1}
\nonumber \\
& \times \left(\frac{\rm [CO/H_2]}{10^{-4}}\right)^{-1}
\frac{\rm M_{\odot }}{\rm K\,km\,s^{-1}\,pc^2},
\end{eqnarray}

\noindent with $T_{\circ} = E_{\rm u}/k_{\rm B}\sim 5.5$\,\K,
$J(T)=T_{\circ} (e^{T_{\circ}/T}-1)^{-1}$, $T_{\rm bg}=(1+z)\,T_{\rm
  CMB}= 9.52$\,\K\ (the temperature of the cosmic microwave background
at $z =2.5$), $g_1=3$ (the degeneracy of level $n=1$), $Z\sim
2\,T_{\rm k}/T_{\circ}$ (the partition function), and where $[{\rm
  CO/H_2}]=10^{-4}$ is the CO abundance in typical molecular clouds
(or a Solar-metallicity environment -- \citealt{bs96}). Note that
$X_{\rm CO}$ is sometimes also known as $\alpha$ and that we shall
ignore its cumbersome units hereafter. The line luminosity, $L'_{\rm
  CO1-0}$, is the velocity- and area-integrated brightness temperature
in the source reference frame, $L'_{\rm \rm CO1-0}= \int_{\Delta V}
\int_{A_{\rm s}} T_{\rm b}\,dA\,dV$, with units
\K\,km\,s$^{-1}$\,pc$^2$ \citep[e.g.][]{solomon97}:

\begin{equation}
L'_{\rm CO1-0} = \frac{3.25\times 10^7}{(1+z)}
\left(\frac{D_{\rm lum}}{\nu_{\rm CO}}\right)^2 
\int_{V}\left[ S_{\rm CO1-0}\,dV\right] ,
\end{equation}

\noindent where $D_{\rm lum}$ is the luminosity distance in Mpc,
$\nu_{\rm CO}$ is the rest-frame frequency (in GHz) of the $^{12}$CO
$J\!=\!1\!-\!0$ transition and the velocity-integrated flux density is
in units of Jy\,km\,s$^{-1}$.

Making no assumption about the state of the molecular gas, we set the
minimum temperature range to $T_{\rm kin}\sim 15-20$\,\K, which yields
$\langle X^{\rm thin}_{\rm CO}\rangle \sim 0.45$, and this can be used
to compute the minimum plausible molecular gas mass in each SMG,
$M_{\rm min}$ (for comparison, a temperature range of $T_{\rm kin}\sim
40-50$\,\K, typical of SF gas, has $\langle X^{\rm thin}_{\rm
  CO}\rangle \sim 0.65$).  We determine $M_{\rm max}$ by adopting
$X_{\rm CO}=5$ (see \S\ref{xco}). Note that the contribution from
Helium is already included.

Setting aside issues regarding the appropriate value of $X_{\rm CO}$,
it is worth mentioning that the use of CO $J\!=\!1\!-\!0$ to estimate
molecular gas masses in SMGs, rather than $J\!=\!3\!-\!2$ or
higher-$J$ lines, yields $\sim$2$\times$ higher masses, simply because
we have observed that $r_{\rm 3-2/1-0}$ is significantly below
unity. If, for a moment, we adopt the so-called `ULIRG-appropriate'
value of $X_{\rm CO}$, which is around 0.8 and has been used widely to
estimate molecular gas masses in SMGs \citep[e.g.][]{greve05,
  tacconi06}, this would yield a median molecular gas mass of $(6.1\pm
2.2)\times 10^{10}$\,M$_{\odot}$ in these systems.  We report the
plausible range of molecular gas masses in Table\,\ref{tab:masses} and
now move on to discuss how best to assess the state of the molecular
gas, and how we might better determine its mass, e.g.\ locally, the
key to choosing an appropriate $X_{\rm CO}$ for ULIRGs came from
dynamical constraints \citep{solomon97}, as we discuss in
\S\ref{dyncon}.

\section{Discussion}
\label{discussion}

\subsection{Physical conditions of the molecular gas}
\label{xco}

Two uncertainties plague us when we employ CO transitions to trace the
mass of molecular gas in high-redshift galaxies, namely: 1) the
assumed \Tb\ ratio, if transitions other than $J\!=\!1\!-\!0$ are
used, and 2) the $X_{\rm CO}$ factor, which must be appropriate for
the average excitation and kinematic state of the molecular gas. The
first of these has been removed by our observation of the $^{12}$CO
$J\!=\!1\!-\!0$ transition and we now focus our discussion on the
second.

In practice, $^{12}$CO $J\!=\!1\!-\!0$ emission in giant molecular
clouds (GMCs) can be optically thick, and $M({\rm H_2}) = X_{\rm CO}
\, L'_{\rm CO1-0}$ is estimated by adopting $X_{\rm CO}\sim 5$. The
latter value of $X_{\rm CO}$ was obtained from GMC studies in
quiescent environments \citep[e.g.][]{solomon87, sb91} and is robust
(to within $\sim$2$\times$) even in warm and dense SF molecular gas,
providing that the molecular gas remains reducible to ensembles of
self-gravitating units in all environments (see \citealt{dickman86}
for the appropriate formalism).  The failure of the latter assumption
results in $X_{\rm CO} \sim 0.3$--1.3 in the extreme interstellar
medium (ISM) of ULIRGs \citep{ds98}, as the molecular gas reservoir
can form a continuous medium rather than an ensemble of virialised gas
clumps.  In the literature, $X_{\rm CO} = 0.8$ is widely adopted for
SMGs, as noted earlier, but this carries a significant uncertainty,
even within the ULIRG class, as the original study indicates.  Such
uncertainties can only be exacerbated in SMGs, e.g.\ more extreme
velocity fields would act to lower $X_{\rm CO}$ still further, while
the presence of dominant quantities of dense ($n(\rm
H_2)>10^4$\,cm$^{-3}$) SF gas could increase $X_{\rm CO}$ -- perhaps
even back into the Galactic range.  For these reasons we do not adopt
a single $X_{\rm CO}$ value, but explore its possible range, aided by
the available CO lines, with the $J\!=\!1\!-\!0$ transition now
providing a normalisation to the total molecular gas mass.

Since we want to avoid the assumption that the molecular gas
reservoirs in SMGs are reducible to ensembles of virialised clumps, as
well as being able to incorporate any information about the average
state of the gas (as provided by available $^{12}$CO lines), we adopt
the following expression:

\begin{equation}
X_{\rm CO} = 2.1 \left(\frac{\sqrt{\langle n({\rm H_2})\rangle}}{T_{\rm CO1-0}}\right)
K^{-1} _{\rm vir}\,{\rm M}_{\odot}\,\left({\rm K\,km\,s^{-1}\, pc^2}\right)^{-1}
\end{equation}

\noindent
where $\langle n({\rm H_2})\rangle$ and $T_{\rm CO1-0}$ are the
average density and \Tb\ (in $^{12}$CO $J\!=\!1\!-\!0$) of the
molecular cloud ensemble \citep[e.g.][]{bs96} to be constrained by
radiative transfer models of available $^{12}$CO line ratios. The
expression:

\begin{equation}
K_{\rm vir} = \frac{\left(dV/dr\right)_{\rm obs}}{\left(dV/dr\right)_{\rm vir}}
\sim 1.54 \frac{[{\rm CO/H_2}]}{\sqrt{\alpha}\,\Lambda _{\rm CO}}
\left[\frac{\langle n({\rm H_2})\rangle}{10^3\, {\rm cm^{-3}}}\right]^{-1/2}
\end{equation}

\noindent
corrects $X_{\rm CO}$ for non-virial gas motions, where $\Lambda_{\rm
  CO}=[{\rm CO/H_2}]/(dV/dR)$ in (km\,s$^{-1}$\,pc$^{-1}$)$^{-1}$.
The parameter, $\alpha = 1$--2.5, depends on the assumed density
profile of a typical cloud; here we adopt an average value, $\alpha =
1.75$. Values of $K_{\rm vir}>1$ bring $X_{\rm CO}$ towards the lower
values reported for ULIRGs \citep{solomon97, ds98} while $K_{\rm
  vir}\ll 1$ is used to exclude dynamically-unattainable kinematic gas
states from the range of radiative transfer modeling solutions.

We can then use $^{12}$CO line ratios to constrain the average
conditions of the molecular gas for the SMGs in our sample, and
equation (3) to obtain the corresponding $X_{\rm CO}$ values.

For the system with the highest level of global $^{12}$CO excitation,
SMM\,J163650, our large-velocity gradient (LVG) modeling of
$r_{3-2/1-0}$ and $r_{6-5/3-2}$ yields two plausible solutions,
namely: a) $T_{\rm kin}\sim 20$--30\,\K\ with $n({\rm H_2})\sim
10^4$--$10^5$\,cm$^{-3}$ and $K_{\rm vir}\sim 4$--11, which
corresponds to $\langle X_{\rm CO}\rangle \sim 4.5$ and b) $T_{\rm
  kin}\sim 90$--110\,\K\ with $n({\rm H_2})\sim 10^3$\,cm$^{-3}$, and
$K_{\rm vir}\sim 1$, with $\langle X_{\rm CO}\rangle \sim 1$ for the
corresponding gas phases. The system with the lowest excitation,
SMM\,J123707-NE, has $r_{3-2/1-0}\sim 0.4$, which is Milky Way-like
and is also compatible with two ranges of LVG solutions: a) $T_{\rm
  kin}=15$\,\K, $n({\rm H_2})=3\times 10^2$\,cm$^{-3}$, $K_{\rm
  vir}\sim 1$ with $X_{\rm CO}=10$, and b) $T_{\rm kin}\sim 25$ or
45--100\,\K, $n({\rm H_2})=(1$--$3)\times 10^2$\,cm$^{-3}$, $K_{\rm
  vir}\sim 4$--10 (and even up to 30) with $\langle X_{\rm CO}\rangle
\sim 0.8$.

Thus $X_{\rm CO}\sim 5$--10 is certainly compatible with the global
$^{12}$CO line excitation observed in SMGs, as are values as low as
$X_{\rm CO}\sim 0.4$--1.  This dual range of LVG solutions and
corresponding $X_{\rm CO}$ factors is a general characteristic of such
modeling when constrained by a small number of line ratios.  In our
LVG models we always find more parameter space associated with the
lower rather than the higher values of $X_{\rm CO}$, but this does not
make the latter less probable, nor does it help break the
aforementioned degeneracy (although as we see in the following
section, dynamical arguments favour the former solutions). It is
important to note that the choice between the two ranges of $X_{\rm
  CO}$ that could apply in SMGs is not between quiescent and SF
molecular gas: $X_{\rm CO}\sim 5$--10 can be associated both with
cold, non-SF gas as well as with the warm, dense gas found in SF
regions.  Both components can be distributed widely in IR-luminous
spiral galaxies and their similar $X_{\rm CO}$ factors actually ensure
the robustness of global molecular gas mass estimates in such systems
when using a common (usually Galactic) $X_{\rm CO}$ value
\citep[e.g.][]{ys91}. Low $X_{\rm CO}$ values, on the other hand, are
found in the warm, diffuse, and typically non-self-gravitating gas
which -- for typical spirals -- is confined to their nuclei
\citep[e.g.][]{regan00}, and thus a low $X_{\rm CO}$ applies for only
a small fraction of their total molecular gas mass.  In ULIRGs, low
$X_{\rm CO}$ factors (with a considerable dispersion) apply for all of
their molecular gas as the aforementioned gas phase is expected to be
concomitant with the entire gas reservoir \citep[e.g.][]{aalto95} --
the result of high pressures in the ISM and tidal molecular cloud
disruption in merger environments.

The difference between molecular gas phases with high $X_{\rm CO}$
(quiescent or SF) and those with low $X_{\rm CO}$ (found mostly in SF
environments as a second phase dominating the low-$J$ $^{12}$CO
emission) is mostly due to the optical depth of the $^{12}$CO
$J\!=\!1\!-\!0$ line.  This is large for the former ($\tau_{1-0}\ga
5$) and typically small for the latter ($\tau_{1-0}\sim 0.5$--2) as a
result of higher temperatures, lower densities and/or velocity
gradients, with $K_{\rm vir}\ga 5$.  The proximity of some ULIRGs has
allowed several studies to break the degeneracies
\citep[e.g.][]{lisenfeld00, hr06}. When using low-$J$ $^{12}$CO lines
this is possible with high-resolution imaging that resolves the gas
disks in ULIRGs \citep{ds98}, while $^{13}$CO and multi-$J$ $^{12}$CO
line measurements \citep[e.g.][]{mao00, weiss01} that allow the mean
optical depth of the $^{12}$CO $J\!=\!1\!-\!0$ line to be determined
can also achieve this -- see \S\ref{degeneracies}.  Finally, detailed
models of CO line emission from turbulent gas clouds reveal that CO
$J\!=\!3\!-\!2$ and higher-$J$ lines are very sensitive to the
presence of dense, self-gravitating cores; these lines are
significantly under-luminous with respect to CO $J\!=\!1\!-\!0$ for
diffuse, non-self-gravitating, turbulent gas \citep{ossenkopf02}.  The
resulting non-linear dependence of high-$J$ ratios such as CO
$J\!=\!7\!-\!6/J\!=\!3\!-\!2$ (available for several SMGs) on the mean
density of such a diffuse phase \citep[see Figure~9 of][]{ossenkopf02}
makes them unsuitable for constraining an unobserved CO
$J\!=\!1\!-\!0$ line luminosity, especially when such a phase exists
alongside dense, self-gravitating, SF gas, as is the case with ULIRGs.

%
%
\begin{table*}
  \centering
  \caption{SMG derived properties.\label{tab:masses}}
  \begin{tabular}{lccccccc}
    \hline 
    \multicolumn{1}{l}{Name} &
    \multicolumn{1}{c}{$L'_{\rm CO1-0}$} &
    \multicolumn{1}{c}{$M^a_{\rm min}, M^b_{\rm max}$} &
    \multicolumn{1}{c}{$m_{\rm CQ/SF}$} &
    \multicolumn{1}{c}{$L_{\rm IR}$} &
    \multicolumn{1}{c}{$M^c_{\rm best}$ range} &
    \multicolumn{1}{c}{$M^d_{\ast}$} &
    \multicolumn{1}{c}{$M_{\rm dyn}$} \\
    \multicolumn{1}{l}{} &
    \multicolumn{1}{c}{($10^{10}$\,{\sc k}\,km\,s$^{-1}$\,pc$^2$)} &
    \multicolumn{1}{c}{(10$^{10}$ M$_{\odot}$)} &
    &
    \multicolumn{1}{c}{($10^{12}$\,L$_{\odot}$)} &
    \multicolumn{1}{c}{($10^{10}$\,M$_{\odot}$)} &
    \multicolumn{1}{c}{($10^{10}$\,M$_{\odot}$)} &
    \multicolumn{1}{c}{($10^{10}$\,M$_{\odot}$)} \\
    \hline
SMM\,J123549   &$7.6\pm1.0$ &3.4, 38&1.3 &$5.5\pm 1.2$&2.5--7.5 &$21\pm 6$&$23\pm 4$\\
SMM\,J123707-NE&$2.7\pm0.6$ &1.2, 13&5.0 &$4.1\pm 0.8$&4.9--14.7&$8\pm 2$ &$18\pm 6$\\
SMM\,J123707-SW&$3.5\pm0.6$ &1.6, 18&1.4 &$3.1\pm 0.6$&1.5--3.5 &$18\pm 3$&$\leq 4$\\
SMM\,J163650   &$9.3\pm1.1$ &4.2, 47&0.33&$7.8\pm 1.5$&2.1--6.3 &$14\pm 4$&$41\pm 8$\\
SMM\,J163658   &$10.6\pm2.0$&4.9, 54&1.5 &$6.4\pm 0.9$&3.2--9.6 &$13\pm 3$&$58\pm 4$\\
    \hline
  \end{tabular}

  {\small
    Notes: $a)$ adopting $X_{\rm CO}=0.45$ -- see \S\ref{masses} and \S\ref{xco};
    $b)$ adopting $X_{\rm CO}=5$ -- see \S\ref{masses} and \S\ref{xco};\\
    $c)$ assuming SFE = 500\,L$_{\odot}$\,M$^{-1}_{\odot}$, and plausibly up to 3$\times$ lower -- see \S\ref{twophases};
    $d)$ from Hainline et al.\ (2010).}
\end{table*}

\subsection{Dynamical constraints on $X_{\rm CO}$ and
            gas mass}
\label{dyncon}

Table~\ref{tab:masses} lists the stellar masses for our target
galaxies, derived using Bruzual \& Charlot stellar population
models\footnotemark[3] \citep{hainline10}.  We caution that these mass
estimates are systematically uncertain due to the potentially complex
star-formation histories and dust obscuration within SMGs, as
discussed by \citet{hainline10}.  We also use the measured {\sc fwhm}
velocity of the $^{12}$CO $J\!=\!1\!-\!0$ lines and the observed
semi-major axes to determine dynamical masses, correcting for
inclination using $\left\langle\sin^2 i \right\rangle=2/3$
\citep[following][]{tacconi08}. Using $M_{\rm dyn}=2.1\sigma_{\rm
  1-0}^2 R/G$ we derive a median dynamical mass within $R\sim 7$\,kpc
of $(2.3\pm 1.4)\times 10^{11}$\,M$_{\odot}$, consistent with the
masses of SMGs from their resolved dynamics \citep{swinbank06} and
roughly 6$\times$ more massive than UV-selected galaxies at this epoch
\citep{erb06}, calculated in the same manner.

\footnotetext[3]{For SMM\,J123707-NE, we scale the mass given by
  \citet{hainline10} for SMM\,J123707-SW based on the relative
  5.8-$\mu$m fluxes of the components.}

Combining our minimum estimate of the molecular gas mass, $\langle
M_{\rm min} \rangle=(3.4\pm 1.2)\times 10^{10}$\,M$_{\odot}$ and the
stellar masses from \citet{hainline10}, $\langle M_{\ast}
\rangle=(1.4\pm 0.3)\times 10^{11}$\,M$_{\odot}$, we derive a total
baryonic mass, $\langle M_{\rm baryon} \rangle = (1.8\pm 0.1)\times
10^{11}$\,M$_{\odot}$ for our five SMGs, comparable with their median
dynamical mass.  We also determine a median gas mass fraction of
$\langle M_{\rm min}\rangle/\langle M_{\rm baryon}\rangle = 0.14\pm
0.02$, a median gas-to-stars mass ratio of $\langle M_{\rm
  min}\rangle/\langle M_{\rm \ast}\rangle = 0.16\pm 0.02$ and a total
baryonic fraction, $f_{\rm baryon} = \langle M_{\rm
  baryon}\rangle/\langle M_{\rm dyn}\rangle = 0.51\pm 0.10$. Adopting
the Milky Way value for $X_{\rm CO}$ would increase the first two of
these fractions by a factor of $\sim$10$\times$ and would result in
$f_{\rm baryon} \sim 1.5$. We note that our median baryonic mass is
about 50 per cent higher than $L^{\ast}$ at the present day
\citep{cole01} suggesting that SMGs are indeed the progenitors of
massive galaxies.

The median baryonic mass does not exceed the median dynamical mass for
$X_{\rm CO} \ls 3$.  However, we note that the mass-to-light ratios
used to derive the stellar masses are not consistent with the
star-formation histories implied by the observed gas mass, the current
SFRs and the expectation that we are (on average) seeing the SMGs
mid-way through their burst phase. If we require that SMGs are seen
half-way through their burst, that the current SFR is 50-per-cent
efficient and is sustained for the duration of the burst, then we can
use the observed dynamical masses, $^{12}$CO $J\!=\!1\!-\!0$ and
rest-frame $H$-band luminosities to derive self-consistent constraints
on $X_{\rm CO}$, the mass of the stellar population prior to the
burst, and the duration of the burst. We use a constant-SFR model from
{\sc starburst99} \citep{leitherer99} to determine the mass-to-light
ratio in the $H$-band of the burst (which is obscured by a foreground
dust screen with extinction, $A_V$) and a pre-existing, unobscured
1-Gyr stellar population whose combined luminosity is required to
reproduce the observed $H$-band flux. A disk-like dynamical model for
the molecular gas reservoir in the SMGs yields $X_{\rm CO} < 2$, an
expected burst lifetime of $\ls$150\,Myr, a gas fraction of $\ls$65
per cent, and moderate extinction for the burst, $A_V\gs 7$.  In this
scenario $X_{\rm CO}=0.8$ is recovered for a model with a gas fraction
of 25 per cent, a pre-existing stellar mass of $1.4\times
10^{11}$\,M$_\odot$ and a burst duration of 50\,Myr with $A_V\sim 20$,
which has added 20 per cent to the stellar mass of the galaxy.
However, allowing a wider range of dynamical models (e.g.\ a
virialised sphere) removes this constraint, allowing solutions as high
as $X_{\rm CO}\sim 5$, and so we conclude that it is impossible with
current information to reliably constrain $X_{\rm CO}$ from dynamical
limits on SMG samples.  Nevertheless, it is clear that values of
$X_{\rm CO}\gs 5$ are disfavoured.

\subsection{The molecular gas in SMGs: two phases}
\label{twophases}

Larger values of $X_{\rm CO}$ have the potential to radically re-shape
our view of the structure, gas-consumption timescales and evolutionary
state of SMGs.  From their observed $^{12}$CO line ratios it is
obvious that they are not dominated by SF, dense and warm gas where
$r_{3-2/1-0}\sim 1$ and $r_{6-5/3-2}\sim 0.8$--1, as measured for
Orion\,A-type clouds and the central SF regions of nearby starbursts
such as M\,82 and NGC\,253 \citep[e.g.][]{wild92, bradford03}.  For
SMGs, the average value of $r_{3-2/1-0}$ is lower, with some
individual sources reaching Milky Way-like values, $r_{3-2/1-0}\sim
0.4$ and $r_{6-5/3-2}\sim 0.23$--0.35, well below that of SF gas.

Recalling our discussion in \S\ref{xco}, such low $^{12}$CO line
ratios do not necessarily imply the presence of cold, quiescent gas
such as that typifying quiescent GMCs in the Milky Way. Diffuse, warm
and highly non-virial gas, concomitant with the SF phase in SMGs, can
also suppress their global $^{12}$CO ratios.  Such a phase is
responsible for the low average $r_{3-2/1-0} \sim 0.66$ in the nuclei
of nearby IR-luminous galaxies \citep[e.g.][]{yao03, leech10} where
its presence has been known for some time \citep{regan00}.
$r_{3-2/1-0}$ then drops from its intrinsic starburst value, $\sim$1,
as the diffuse phase contributes extra $^{12}$CO $J=1$-0 emission,
while the $J\!=\!3\!-\!2$ and higher-$J$ lines remain dominated by the
SF gas.  $^{12}$CO line ratios normalised to the $J\!=\!1\!-\!0$ line
luminosity (e.g.\ $r_{3-2/1-0}$) would then be lower than those
intrinsic to SF gas, while ratios involving only high-$J$ lines (e.g.\
$r_{6-5/3-2}$) would be high, being dominated by the SF gas phase.
{\it This is not the case for any of the SMGs in our sample}
(Table~\ref{tab:properties}).  Moreover, LVG modeling finds no average
gas state that can adequately reproduce both $r_{3-2/1-0}$ and
$r_{6-5/3-2}$ for any of the SMGs in our sample, with fits of only the
high-$J$ CO ratios yielding $r_{3-2/1-0}\sim 0.9$--1.2, which is much
higher than observed.  Thus, while a broad two-phase differentiation
of the molecular gas is certainly apparent in SMGs, {\it it is not of
  the type observed in local starburst nuclei and ULIRGs} (with its
associated low $X_{\rm CO}$).  Significant masses of non-SF molecular
gas remain the only viable alternative for the low $r_{3-2/1-0}$ and
$r_{6-5/3-2}$ observed in SMGs.  This is further corroborated by the
fact that the systems with the largest spatial extent of the $^{12}$CO
$J\!=\!1\!-\!0$ relative to $J\!=\!3\!-\!2$ emission show the lowest
$r_{3-2/1-0}$ ratios (see Fig.~\ref{fig:both} and
Table~\ref{tab:properties}).

If we assume conservatively that $X_{\rm CO}$(SF)$\,\sim X_{\rm
  CO}{\rm (quiescent)}\sim 5$ for both the SF and quiescent gas in
SMGs (in practice $X_{\rm CO}$ for the SF gas can be somewhat lower),
then we arrive at robust upper limits for their total molecular gas
masses, $M_{\rm max}$, as described in \S\ref{masses} and listed in
Table~\ref{tab:masses}. Alternatively, we can use $r_{3-2/1-0}$ to
determine the relative fractions of cold, quiescent (CQ) and warm, SF
molecular gas $m_{\rm CQ/SF}=M_{\rm CQ}/M_{\rm SF}$ from:

\begin{equation}
m_{\rm CQ/SF}=\frac{r^{\rm (SF)}_{3-2/1-0}-r_{3-2/1-0}}
{r_{3-2/1-0}-r^{\rm (CQ)}_{3-2/1-0}},
\end{equation}

\noindent
where $r^{\rm (SF)}_{3-2/1-0}\sim 0.9$ and $r^{\rm (CQ)}_{3-2/1-0}\sim
0.3$ are set as typical intrinsic ratios of these two gas phases.  The
total molecular gas mass, $M_{\rm best}=M_{\rm SF}(1+m_{\rm CQ/SF})$,
can then be deduced using $L_{\rm IR}$ (\S\ref{lir}) if we assume that
the SF gas phase in all galaxies has the same intrinsic ${\rm
  SFE}_{\rm max}=L_{\rm IR}/M_{\rm SF}$. Its maximum value,
$\sim$500\,L$_{\odot}$\,M$_{\odot}^{-1}$, is thought to be the result
of an Eddington limit set by photon pressure on dust in the molecular
gas accreted by the SF sites \citep{scoville04, thompson09} while an
almost identical limiting SFE can result from a cosmic-ray-generated
Eddington limit \citep{socrates08}. Such `maximum SFE' values have
been observed in diverse places: individual molecular clouds around OB
star clusters in M\,51, for the total molecular gas reservoir of
Arp\,220, as well as for the HCN-bright gas phase of LIRGs. Here we
adopt the maximum SFE, noting that the true value will likely be
smaller by a factor of up to 3$\times$ since gas accretion towards
sites of star formation may not be spherically symmetric; moreover, CO
emission may also be included from beyond the natal sites of the OB
stars (where the Eddington limit is set), which explains why the
HCN-bright gas phase yields higher values of SFE in ULIRGs, closer to
the maximum value than those associated with the CO-bright gas phase
\citep[e.g.][]{gs04}.

$m_{\rm CQ/SF}$ and $M_{\rm best}$ are listed in
Table~\ref{tab:masses}.  Reassuringly, the maximum globally-averaged
$\langle {\rm SFE} \rangle$ -- estimated using $M_{\rm min}$ (via
$X^{\rm thin}_{\rm CO}$) for the total gas mass and making no
discrimination between SF and non-SF molecular gas -- is around
160--200\,L$_{\odot}$\,M$_{\odot}^{-1}$ for most of our SMGs, with
only SMM\,J163650 approaching the maximum value allowed by our
aforementioned arguments.

The median cold/warm gas mass fraction for the SMGs is 1.4
(Table~\ref{tab:masses}). Indeed, only in one source, SMM\,J163650, do
we find $m_{\rm CQ/SF} < 1$, strong evidence that a significant
fraction of the molecular gas in most SMGs must be cold,
low-excitation gas, taking no part in the starburst.\footnotemark[4]
The large $m_{\rm CQ/SF}$ values, along with evidence for extended
$^{12}$CO $J\!=\!1\!-\!0$ emission \citep[this
study;][]{ivison10leblob, carilli10}, suggests that the main gas
reservoir in these systems is more widely distributed than the
compact, maximal starbursts implied by high-$J$ $^{12}$CO imaging
\citep[e.g.][]{tacconi06}.

We consider that $M_{\rm best}$ provides the most likely range of the
total molecular gas masses of the SMGs in our sample, albeit with a
large uncertainty due to the plausible range of maximum SFE. Using
this new method, we find a median molecular gas mass of $(2.5\pm 0.8)
\times 10^{10}$\,M$_{\odot}$, with a plausible range stretching up to
3$\times$ higher.  If we combine both components in SMM\,J123707, this
becomes $(3.2\pm 0.9)\times 10^{10}$\,M$_{\odot}$.

\footnotetext[4]{Any significant AGN contribution to $L_{\rm IR}$
  would lower our estimates of the warm SF molecular gas mass and
  increase the $m_{\rm CQ/SF}$ gas mass fractions reported in
  Table~\ref{tab:masses}.}.

\subsection{Breaking degeneracies in  excitation and $X_{\rm CO}$ via
 $^{13}$CO observations}
\label{degeneracies}

During 2011--12 the EVLA will offer 2\,GHz of instantaneous bandwidth,
prior to commissioning of the full 8-GHz WIDAR capability.  This will
allow simultaneous\footnotemark[5] imaging of $^{12}$CO and $^{13}$CO
$J\!=\!1\!-\!0$, each having 4$\times$ the velocity coverage utilised
to date.

\footnotetext[5]{The frequency of $^{13}$CO $J\!=\!1\!-\!0$ differs
  from that of $^{12}$CO $J\!=\!1\!-\!0$ by $5.070/(1+z)$\,GHz, or
  $\sim$1.5\,GHz for the SMGs targeted here, though note that
  contiguous placement of sub-band pairs is not mandatory when using
  WIDAR.}

Sensitive $^{13}$CO $J\!=\!1\!-\!0$ imaging of SMGs can break the
degeneracies of the $^{12}$CO spectral-line energy distribution
(SLED), allowing us to determine $X_{\rm CO}$ by: a) determining the
total molecular gas mass via an independent, typically optically thin
line, or b) placing diagnostically powerful lower limits on the
$^{12}$CO/$^{13}$CO $J\!=\!1\!-\!0$ line-intensity ratio, $R_{1-0}$.
For the $^{12}$CO intensities seen in our sample, deep EVLA
observations can detect $^{13}$CO $J\!=\!1\!-\!0$ where $R_{1-0}\sim
3$--6, typical for cold Galactic GMCs, where $^{12}$CO $J\!=\!1\!-\!0$
is optically thick \citep{scoville79, polk88}.  On the other hand, the
ability to set lower limits of $R_{1-0}>10$ can decisively break the
degeneracy described in \S\ref{twophases} in favour of the diffuse
phase with moderate optical depth in $^{12}$CO $J\!=\!1\!-\!0$ and low
$X_{\rm CO}$.  This is because while $R_{1-0}\sim 5$--10 is typical of
Galactic GMCs and quiescent spiral disks \citep[e.g.][]{sakamoto97,
  paglione01}, $R_{1-0}>10$ is found almost exclusively in starbursts
(e.g.\ \citealt*{casoli92}; \citealt{aalto95}) or galactic centres,
where a diffuse, warm, non-self-gravitating phase dominates the
$J\!=\!1\!-\!0$ emission.

For the SMG with the lowest $^{12}$CO excitation in our sample,
SMM\,J123707-NE, our LVG models (for an abundance ratio $\rm
[^{12}CO/^{13}CO]=60$) yield $R_{1-0}\gs 20$ for all solutions
corresponding to diffuse, warm, non-self-gravitating gas, while
$R_{1-0}\sim 9$--10 is found for the dense, cold, self-gravitating
phase.  Similarly, for the average $r_{3-2/1-0}$ of the sample, and
$R_{1-0}=5$, all LVG solutions correspond to virial or only slightly
unbound kinematic states ($K_{\rm vir}\sim 1$--3) with $X_{\rm CO}\sim
3$--6, with the best fit found for $T_{\rm k}\sim 15$\,\K\ and $n({\rm
  H}_2)\sim 10^3$\,cm$^{-3}$ -- conditions typical for quiescent
GMCs. For $r_{3-2/1-0}\sim 0.6$ and $R_{1-0}\sim 15$, all solutions
compatible with the aforementioned \Tb\ ratio have $K_{\rm vir}\gs 5$
and a corresponding $X_{\rm CO}\sim 0.65$--1.2.  For the highest
$^{12}$CO $J\!=\!1\!-\!0$ flux levels observed in our sample, a
$\sim$10$\times$ weaker $^{13}$CO $J\!=\!1\!-\!0$ line can be detected
at $\gs$10$\sigma$ by the fully upgraded EVLA in $\ls$100\,hr.

Finally, we mention the possibility that there may be a significant
optical depth due to dust in compact ULIRGs with very dense gas
($n({\rm H}_2)\ga 10^5$\,cm$^{-3}$), even at short submm wavelengths.
This is suspected for nearby ULIRGs such as Arp\,220, and for some
SMGs \citep{p10b}, and creates an additional degeneracy for the
interpretation of very low (high-$J$)/(low-$J$) $^{12}$CO line ratios
in extreme starbursts.  EVLA observations of the high-density gas
tracer, HCN $J\!=\!1\!-\!0$, at 88.632\,GHz, along with low-$J$
$^{12}$CO transitions and high-$J$ ($J\!=\!6\!-\!5$ and higher)
$^{12}$CO transitions with the Atacama Large Millimetre Array (ALMA),
have the potential to break this degeneracy \citep{papadopoulos10a}.

%
%
\begin{figure}
\centerline{\psfig{figure=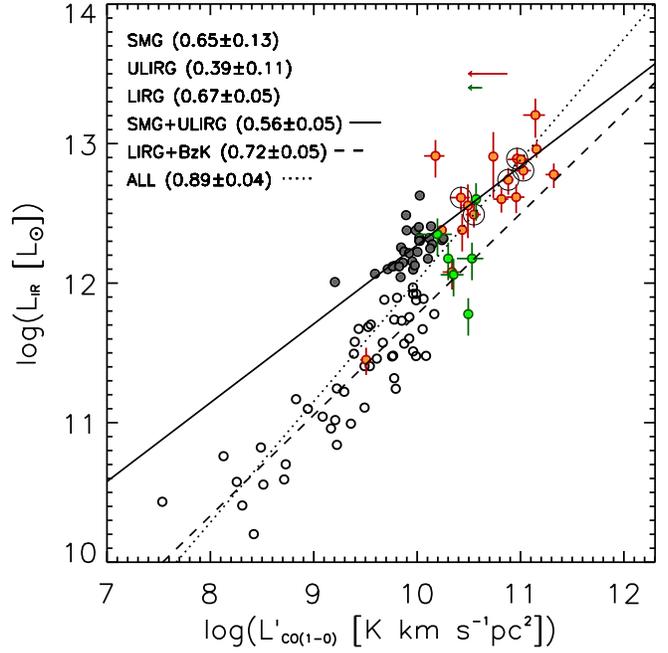,width=3.4in,angle=0}}
\caption{$L_{\rm IR}$ versus $L'_{\rm CO}$ for those SMGs, local
  (U)LIRGs and BzK galaxies (see legend) with robust CO
  $J\!=\!1\!-\!0$ (or $J\!=\!2\!-\!1$) measurements. Trend lines are
  fitted to the ULIRGs plus SMGs (solid) and LIRGs plus BzK galaxies
  (dashed), as well as all the samples combined (dotted). The
  resulting linear slopes are given in parentheses. Also shown are the
  mean corrections that would have been required in $L'_{\rm CO}$ for
  the SMGs and BzK galaxies if their CO luminosities had been
  derived using CO $J\!=\!4\!-\!3$ or $J\!=\!3\!-\!2$, which would
  have resulted in an erroneously steep trend. The measurements from
  this paper are circled.}
\label{fig:sk}
\end{figure}

\subsection{Gas depletion timescales and the evolutionary state of SMGs}
\label{tau}

The gas-consumption timescale, $\tau_{\rm gas}=M({\rm H}_2)/{\rm
  SFR}$, is the time needed for star formation to consume a given
molecular gas reservoir, in the absence of feedback effects, and can
be estimated using the plausible range of values for $M_{\rm SF}$ and
$M_{\rm best}$. Thus $\tau_{\rm gas}({\rm SF}) = M_{\rm SF}/{\rm SFR}$
yields the shortest feasible duration for the observed SMG bursts,
while $M_{\rm best}/{\rm SFR}=\tau_{\rm gas}({\rm SF}) (1+m_{\rm
  CQ/SF})$ is the longest time over which SF can continue if the
colder material can be involved without delays imposed by dynamics or
SF-related feedback.

We find a median gas-consumption timescale, $\langle \tau_{\rm
  gas}({\rm SF})\rangle =12$\,Myr, which lengthens to 28\,Myr if the
colder gas can become involved. This is considerably shorter than
naive expectations for the lifetime of the SMG phase, $\sim$300\,Myr
\citep[e.g.][]{swinbank06}, but these gas-comsumption timescales can
be extended if the SFE is below 500\,L$_{\odot}$\,M$_{\odot}^{-1}$,
perhaps up to 3$\times$ longer.  Moreover, we are neglecting the
effect of feedback, which is also expected to lengthen the total
duration of successive SF episodes in SMGs due to the need to
re-accrete material towards the typically very compact SF regions
($\sim$100--500\,pc) expected in merger-driven, gas-rich starbursts.

\subsection{The rest-frame Schmidt-Kennicutt relation in the distant
            Universe: a comparison to the local Universe}
\label{s-k}

The availability of a significant number of $^{12}$CO $J\!=\!1\!-\!0$
line luminosities in vigorously SF systems at high redshifts presents
an opportunity to examine the Schmidt-Kennicutt (S-K) relation between
their gas reservoirs and SFRs, without resorting to the use of
high-$J$ $^{12}$CO transitions and assumptions about global $^{12}$CO
high-$J$/$J\!=\!1\!-\!0$ line ratios (i.e.\ assumptions about the
average state of their molecular gas reservoirs). The latter can
create artificial offsets from the S-K relation for galaxy populations
where only high-$J$ $^{12}$CO lines have been observed, if an
incorrect $^{12}$CO high-$J$/$J\!=\!1\!-\!0$ ratio is assumed. Such
offsets can also be created by the application of inappropriate
$X_{\rm CO}$ factors in various galaxy classes.  For our present short
discussion we render the S-K relation only as a $L_{\rm IR}-L'_{\rm
  CO}$ relation, postponing our investigation of the physical
relation, ${\rm SFR}-M({\rm H}_2)$, where $X_{\rm CO}$ must be
considered, for a future paper (Greve et al., in preparation).

In its original form, the S-K relation was established via H\,{\sc i}
and $^{12}$CO $J\!=\!1\!-\!0$ measurements of galaxies in the local
Universe \citep{schmidt59, kennicutt89, kennicutt98}. The H\,{\sc i} +
H$_2$ gas surface densities, $\Sigma$(H\,{\sc i}+H$_2$), are related
to the star-formation surface density, $\Sigma_{\star}$, by
$\Sigma_{\star}\propto \left[\Sigma({\rm H\,I+H_2})\right]^{\kappa}$,
where $\kappa \sim 1.4$--1.5. In a recent, comprehensive analysis of
the H\,{\sc i} and H$_2$ distributions in nearby galaxies, on sub-kpc
scales, \citet{bigiel09} found a power-law with slope $N=1.0\pm 0.2$
in regions where the total gas content (H\,{\sc i} + H$_2$) is
dominated by H$_2$. The latter seems to be characterised by gas
surface densities $\gs$9\,M$_{\odot}$\,pc$^{-2}$, indicating a
saturation point for H\,{\sc i}.

Assuming that distant LIRGs are mostly H\,{\sc i}-poor, as they are in
the local Universe, with $\ls$20 per cent of the their total gas in
H\,{\sc i}, we can estimate their total mass using the $^{12}$CO
$J\!=\!1\!-\!0$ line.  Even for the most H\,{\sc i}-rich local LIRGs,
omission of H\,{\sc i} results in a gas-mass under-estimate of
$\ls$2$\times$, leaving $X_{\rm CO}$ as the dominant uncertainty (see
\S\ref{xco}).

That SMGs follow the S-K relation in its basic observational form,
$L_{\rm IR}-L'_{\rm CO}$, was first shown by the $^{12}$CO
$J\!=\!3\!-\!2$ and $4-3$ survey of SMGs by \citet{greve05}. More
recently, CO surveys of gas-rich disk galaxies at $z=1$--2
\citep{daddi08, daddi10, dannerbauer09, genzel10} have shown that they
too obey a S-K-like relation, albeit apparently offset from the SMGs
(lower $L_{\rm IR}/L'_{\rm CO}$). It has been speculated that this
offset reflects two modes of global star formation in galaxies: SMGs
typify those galaxies undergoing major mergers, with intense, highly
efficient bursts of star formation, and BzKs typify galaxies
undergoing a more leisurely rate of star formation, at lower
efficiency.  However, this apparent offset in the S-K relation is
deduced from a comparison of high-$J$ CO observations from SMGs with a
mixture of CO $J\!=\!3\!-\!2$, $J\!=\!2\!-\!1$ and $1\!-\!0$ line
observations of disk galaxies, and the reality of the apparent offset
(and its interpretation as different global star-formation laws for
mergers and disk galaxies) is therefore hampered by biases and
uncertainties in the underlying gas excitation, especially in SMGs.

With an increased sample of SMGs observed in $^{12}$CO $J\!=\!1\!-\!0$
we can make a relatively unbiased comparison between SMGs and the
$z\sim 1$--2 disk population.  In Fig.~\ref{fig:sk} we show $L_{\rm
  IR}$ (\S\ref{lir}) versus $L'_{\rm CO1-0}$, populated with local
LIRGs and ULIRGs from Papadopoulos et al.\ (in preparation) as well as
with those SMGs and BzK galaxies with reliable measurements of
$^{12}$CO $J\!=\!1\!-\!0$ or $J\!=\!2\!-\!1$ (where $J\!=\!1\!-\!0$ is
not available, we use $J\!=\!2\!-\!1$, calculating $L'_{\rm CO1-0}$
via $r_{21}=0.75$).  We see that the SMGs extend the S-K relation to
higher luminosities, with a low dispersion. Fitting linear relations
of the form $\log L_{\rm IR} = \alpha \log L'_{\rm CO1-0} + \beta$, we
determine slopes for the various samples, reporting these in the
labels of Fig.~\ref{fig:sk}. Taking the SMGs alone, we find
$\alpha=0.65\pm 0.13$ and $\beta=5.7\pm 1.4$; local ULIRGs display a
shallower slope, $\alpha=0.39\pm 0.11$ with $\beta=8.4\pm 1.1$, though
this is highly uncertain as they span only a small range in $L_{\rm
  IR}$; for LIRGs we see $\alpha =0.67\pm 0.05$ and $\beta=5.0\pm
0.5$, similar to the SMGs.

Generally, we find slopes that are significantly below unity
($\alpha\sim 0.5$--0.7) for the various samples. This contrasts with
the steeper slopes ($\alpha\sim 1.1$--1.5) reported by other studies
of low- and high-redshift samples \citep[e.g.][]{greve05, daddi10,
  genzel10}. These have typically employed high-$J$ $^{12}$CO line
luminosities for the high-redshift galaxies and $L'_{\rm CO1-0}$ for
the local sources, which may artificially steepen the slope.
\citet{iono09} looked at the S-K relation using $^{12}$CO
$J\!=\!3\!-\!2$ for both low- and high-redshift galaxies, finding
$\alpha=1.08\pm 0.03$, similar to that derived by \citet{gs04} using
HCN $J\!=\!1\!-\!0$.  These two molecular transitions both trace the
dense, warm, SF gas phase. A common slope of unity in the
corresponding S-K relations is to be expected if a near-constant
SFE$_{\rm max}=L_{\rm IR}/M_{\rm SF}$ underlies star formation in all
galaxies, with only the dense gas phase available as fuel, and an
Eddington-type limit (set by photons or cosmic rays) setting SFE$_{\rm
  max}$ (see \S\ref{twophases}).

Taking the SMGs and ULIRGs together -- they both contain extreme SF
environments, after all -- yields $\alpha=0.56\pm 0.05$ and
$\beta=6.6\pm 0.6$, whereas taking the LIRGs and BzK galaxies together
yields $\alpha=0.72\pm 0.05$ and $\beta=4.6\pm 0.5$.  The BzK galaxies
do not stand out dramatically from the other low- and high-redshift
samples. There are only three BzK galaxies with $^{12}$CO
$J\!=\!1\!-\!0$ detections \citep{aravena10}, but the situation is
unchanged when we include the BzK galaxies detected in $^{12}$CO
$J\!=\!2\!-\!1$ from \citet{daddi10}, adopting $r_{2-1/1-0}=0.75$ to
calculate $L'_{\rm CO1-0}$. If, however, we compare the BzKs with the
SMGs, this time using the $^{12}$CO $J\!=\!3\!-\!2$, $4\!-\!3$ and
$5\!-\!4$ lines for the latter (the mean resulting offset in $L'_{\rm
  CO}$ is shown in Fig.~\ref{fig:sk}), then one might argue that SMGs
and ULIRGs populate a different sequence to BzK and spiral galaxies,
as did \citet{daddi10}. Thus an apparent displacement and/or
steepening of the S-K relation between various galaxy populations may
not reflect a true difference in their respective SF modes, but rather
the strong excitation biases produced by using molecular lines with
different excitation requirements. Given these biases, the relatively
small galaxy samples, and their scatter, the evidence for different
S-K relations between different galaxy populations is weak, as we
shall argue in a forthcoming paper (Greve et al., in preparation).

Adding a last cautionary note, \citet{pp10} have shown that while
typical present-day galaxies quickly settle into S-K-type relations,
this may not be the case for gas-rich, metal-poor systems in the early
Universe. Such galaxies can spend sustained periods with their SFR
significantly below or above that expected from the local S-K
relation. If the same gas-rich, metal-poor galaxy can deviate strongly
from the S-K relation during the course of its evolution, then the use
of the S-K relation as a tool for differentiating between different SF
modes must be re-evaluated. Indeed, given the dynamic and
non-equilibrium ISM in strongly evolving, gas-rich systems (where the
global mass fractions of the various gas phases, e.g.\ $M_{\rm
  SF}({\rm H}_2)/M_{\rm total}$ and $M($H\,{\sc i}$)/M({\rm H}_2)$,
are strongly time-dependent) there could be a simple, underlying S-K
relation of ${\rm SFR}\propto M_{\rm SF}({\rm H}_2)$, while the
different exponents recovered in various samples are artifacts of
strongly evolving $M_{\rm SF}({\rm H}_2)/M_{\rm total}$ fractions,
further compounded by the choice of observed molecular lines so with
different excitation requirements for high-redshift systems.

\section{Future prospects}

\subsection{Blind surveys}

\citet*{blain04} present the number of blind CO $J\!=\!1\!-\!0$ line
detections expected with 4-GHz-wide observing bands, based on
\citet{blain02}, updating the work of \citet*{carilli02} to reflect
the SMG redshift distribution of \citet{chapman05}. Their 30--34-GHz
band is most easily compared with our $z\sim2.4$ SMGs: scaling to our
236-MHz of instantaneous bandwidth, they predict a source density of
$\sim$2.4\,deg$^{-2}$ at the flux levels (\S\ref{observations}) to
which we are sensitive, $\gs 2\times 10^{-22}$\,W\,m$^{-2}$ or
$\gs$0.18\,Jy\,km\,s$^{-1}$.

It is no great surprise, therefore, that we find no robust
($>$5-$\sigma$) detections within the four fields of our pilot survey,
each with a $\sim$70-arcsec {\sc fwhm} primary beam: sensitivities
ranging from $\sim$0.1--0.2\,Jy\,km\,s$^{-1}$, covering
$\sim$10$^{-3}$\,deg$^2$.

In \citeauthor{blain04}, the galaxies responsible for $J\!=\!1\!-\!0$
emission at or above the level to which we are sensitive are extremely
luminous -- they have higher $L_{\rm IR}$ than our SMGs. This suggests
the predictions may be rather pessimistic; of course, we might also
expect to see an over-density of sources since our fields are centred
on relatively massive galaxies \citep[e.g.][]{stevens03}.

Based on the predictions of \citeauthor{blain04}, a blind survey
utilising WIDAR's full 8-GHz bandwidth, would need to cover
$\sim$100$\times$ our current area, to a similar depth, to provide a
significant number of robust detections and hence a useful test of the
predictions. Our survey suggests these requirements can be relaxed
significantly; it will be interesting to see whether the
$\sim$2$\times$ deeper observations planned for our targets using
EVLA's C configuration provide evidence to support our suggestion.

\subsection{The study of cold, quiescent molecular gas at high
  redshift}

All current molecular-lines studies of distant starbursts, including
our own, involve samples selected via large rest-frame far-IR
luminosities and luminous high-$J$ $^{12}$CO lines ($J\!=\!3\!-\!2$
and higher).  It is thus possible that SMGs with low $r_{\rm 3-2/1-0}$
ratios remain mostly undetected by the $^{12}$CO $J\!=\!3\!-\!2$
surveys and were thereby excluded from our study, as noted in
\S\ref{tbratios}.

The fact that much of the molecular gas in SF systems may not be
participating in the starbursts has been well-established by studies
of local LIRGs, where sensitive low-$J$ $^{12}$CO imaging
\citep[e.g.][]{weiss04} or submm continuum imaging of dust
\citep[e.g.][]{papadopoulos99, dunne01, thomas02} could {\it
  spatially} disentangle the cold and extended molecular gas and/or
its associated dust from the compact starbursts, which otherwise
dominate their global SEDs and their molecular SLEDs.

It has only recently become possible to attempt a similar spatial
separation of ISM components in high-redshift systems, via tracers
that can remain luminous in the cold, low-excitation ISM -- e.g.\
$^{12}$CO $J\!=\!1\!-\!0$ and submm continuum emission from dust. This
explains the previous lack of evidence for extended, low-excitation
molecular gas around SMGs, but also point to several obvious routes
forward, via EVLA and ALMA.

Less well known is the prospect of utilising the two fine-structure
lines of neutral carbon, $^3P_1$$\rightarrow$$^3P_0$ at 492.160\,GHz
and $^3P_2$$\rightarrow$$^3P_1$ at 809.343\,GHz to trace molecular gas
and dynamical mass \citep{papadopoulos04, pg04, weiss05ci}, exploiting
the full concomitance of C and CO in molecular clouds \citep{keene96},
the simple, three-level partition function, the low optical depth and
modest excitation requirements.  Moreover, unlike the luminous
[C\,{\sc ii}] fine-structure line at 1.9\,THz, the neutral carbon
lines are not subject to contamination by atomic or ionised gas,
tracing solely the molecular gas.  The [C\,{\sc i}] $J\!=\!1\!-\!0$
line remains luminous even for UV-shielded, cold ($T_{\rm k}\sim
15$\,\K) molecular gas \citep[e.g.][]{oka01} where the [C\,{\sc ii}]
line luminosity is negligable because all carbon is neutral and
$T_{\rm k}\ll \Delta E_{\rm u}$([C\,{\sc ii}])$/k_{\rm B}\sim
92$\,\K. Finally, the [C\,{\sc i}] lines are accessible in the most
sensitive bands (3--7, or 84--373\,GHz) of ALMA for a much wider
redshift range (and range of look-back times) than [C\,{\sc ii}] --
$z\sim 0.3$--4.9 for $J\!=\!1\!-\!0$ and $z\sim 1.2$--8.6 for
$J\!=\!2\!-\!1$ versus $z\gs 4$ for [C\,{\sc ii}].
 
The [C\,{\sc i}] $^3P_1 \rightarrow\ {^3}P_0$ line, in particular,
remains well-excited in quiescent GMCs ($E_{\rm u}/k_{\rm B}\sim
24$\,\K, $n_{\rm crit}\sim 600$\,cm$^{-3}$ for low $T_{\rm k}$), while
maintaining a favourable $K$-correction with respect to $^{12}$CO
$J\!=\!1\!-\!0$ ($S_{\rm C\,I}/S_{\rm CO}\sim 2.75$--5.5 for $T_{\rm
  b}$([C\,{\sc i}])$/T_{\rm b}$(CO)$\sim 0.15-0.30$).  The $^3P_2
\rightarrow\ {^3}P_1$ line ($E_{\rm u}/k_{\rm B}\sim 63$\,\K, $n_{\rm
  crit}\sim 965$\,cm$^{-3}$) also maintains a $K$-correction advantage
with respect to both $^{12}$CO $J\!=\!1\!-\!0$ and $J\!=\!7\!-\!6$
(the CO line closest in rest-frame frequency), while detecting the
pair yields an excellent thermometer for molecular gas (both lines are
available in bands 3--7 of ALMA for $z\sim 1.2$--4.9).  Thus [C\,{\sc
  i}] $J\!=\!1\!-\!0$, $J\!=\!2\!-\!1$ imaging with ALMA and CO
$J\!=\!1\!-\!0$ imaging with the completed EVLA will constitute the
most powerful tools for making inventories of the global molecular gas
and dynamical gas mass in distant galaxies, unbiased by the excitation
state of the molecular gas and the extent of their star-formation
regions.

\section{Conclusions}

We report the results of a pilot study with the EVLA of $^{12}$CO
$J\!=\!1\!-\!0$ emission from a small sample of well-studied SMGs at
$z= 2.2$--2.5, previously detected in $^{12}$CO $J\!=\!3\!-\!2$ using
PdBI.

Using the EVLA's most compact configuration we detect strong, broad
($\sim$1,000\,\kms\ {\sc fwzi}) line emission from all of our targets
-- coincident in position and velocity with their $J\!=\!3\!-\!2$
emission.

The median line width ratio, $\sigma_{1-0}/\sigma_{3-2}=1.15\pm 0.06$,
suggests that the $J\!=\!1\!-\!0$ emission is more spatially extended
than the $J\!=\!3\!-\!2$ emission, a situation confirmed by our maps
which reveal velocity structure in several cases and typical sizes of
$\sim$16\,kpc {\sc fwhm}. With the current spatial resolution we are
unable to determine whether observed gas motions are well ordered, but
we find no evidence of large-scale flows of cold gas.

We find a median \Tb\ ratio of $r_{3-2/1-0}=0.55\pm0.05$, consistent
with local galaxies with $L_{\rm IR}>10^{11}$\,L$_{\odot}$, noting
that our value may be biased high because of the $J\!=\!3\!-\!2$-based
sample selection. Including five systems with similar luminosities
from the literature we find a median of $r_{3-2/1-0}=0.58\pm 0.05$ and
see no evidence for measureable intrinsic scatter within the sample.

Using the observed $^{12}$CO $J\!=\!1\!-\!0$ line emission, naive
estimates of the molecular gas masses are around 2$\times$ higher than
previous estimates based on $^{12}$CO $J\!=\!3\!-\!2$ with
$r_{3-2/1-0}=1$.

We also estimate molecular gas masses using the $^{12}$CO
$J\!=\!1\!-\!0$ line and the observed global \Tb\ ratios, assuming
standard underlying \Tb\ ratios for the non-SF and SF gas phases as
well as a common SFE for the latter in all systems, i.e.\ without
calling upon $X_{\rm CO}$. Using this new method, we find a median
molecular gas mass of $(2.5\pm 0.8) \times 10^{10}$\,M$_{\odot}$, with
a plausible range stretching up to 3$\times$ higher. Even higher
masses cannot be ruled out, but are not favoured by dynamical
constraints: the median dynamical mass within $R\sim 7$\,kpc for our
sample, $(2.3\pm1.4) \times 10^{11}$\,M$_{\odot}$, $\sim$6$\times$
more massive than UV-selected galaxies at this epoch.

We find a median gas-consumption timescale, $\langle \tau_{\rm
  gas}({\rm SF})\rangle =12$\,Myr, or 28\,Myr if the colder gas can
become involved. This is shorter than naive expectations for the
lifetime of the SMG phase, but these timescales can be longer if the
SFE is below 500\,L$_{\odot}$\,M$_{\odot}^{-1}$, and we neglect the
effect of feedback.

We examine the S-K relation in $L_{\rm IR}-L'_{\rm CO}$ for all the
distant galaxy populations for which CO $J\!=\!1\!-\!0$ or
$J\!=\!2\!-\!1$ data are available, finding small systematic
differences between populations. These have previously been
interpreted as evidence for different modes of star formation, but we
argue that these differences are to be expected, given the still
considerable uncertainties, certainly when considering the probable
excitation biases due to the molecular lines used, and the possibility
of sustained S-K offsets during the evolution of individual, gas-rich
systems.

We discuss the degeneracies surrounding molecular gas mass estimates,
the possibilities for breaking them, and the future prospects for
imaging and studying cold, quiescent molecular gas at high redshift.

We note in ending that if SMGs (and other high-redshift starbursts)
are as extended as our observations suggest (up to 20\,kpc) then even
the shortest possible dish spacings of the EVLA are not well-matched
to their sizes.  However, the smaller ALMA dishes (especially those in
the ALMA Compact Array), if fitted with Band-1 receivers, or GBT, are
ideal for studying the critical $^{12}$CO $J\!=\!1\!-\!0$ emission
from these galaxies.

\section*{Acknowledgements}

We would like to express our immense gratitude to the EVLA
commissioning team and all those that have helped to create this
remarkable facility.  We thank Andy Harris, Andrew Baker and Mark
Swinbank for useful discussions, an anonymous referee for comments
that improved our paper significantly, and Mrs Katerina Papadopoulos
for her visit to Pasadena, and for stopping at two.  IRS acknowledges
support from STFC.

\bibliographystyle{mnras}
\bibliography{co10}

\bsp

\end{document}